\documentclass[12pt]{report}
\begin{document}

\begin{titlepage}
\title{Problems in Modern High Performance Parallel I\slash O Systems}
\author{Robert L Cloud\thanks{rcloud@uab.edu}\\The University of Alabama at Birmingham\\Department of Mechanical Engineering}
\date{\today}
\end{titlepage}
\maketitle

\pagebreak 

\section{Abstract}
\label{abstract}

In the past couple of decades, the computational abilities of supercomputers have increased tremendously. Leadership scale supercomputers now are capable of petaflops. Likewise, the problem size targeted by applications running on such computers has also scaled. These large applications have I\slash O throughput requirements on the order of tens of gigabytes per second. For a variety of reasons, the I\slash O subsystems of such computers have not kept pace with the computational increases, and the time required for I\slash O in an application has become one of the dominant bottlenecks. Also troublesome is the fact that scientific applications do not attain near the peak theoretical bandwidth of the I\slash O subsystems. In addressing the two prior issues, one must also question the nature of the data itself; one can ask whether contemporary practices of data dumping and analysis are optimal and whether they will continue to be applicable as computers continue to scale. These three topics, the I\slash O subsystem, the nature of scientific data output, and future possible optimizations are discussed in this report. 

\pagebreak 

\chapter{Introduction}
\label{introduction}

\section{Static I\slash O Performance and Rapidly Scaling Computation}
\label{staticioperformanceandrapidlyscalingcomputation}

A 2005 paper from Lawrence Livermore National Labs \cite{Hedges2005} states that ASC calibre supercomputers, i.e. leadership class machines capable of running the most demanding applications, require I\slash O bandwidth of 1 GB\slash sec per teraflop of computing capability. This rule would enable a 100 teraflop supercomputer to perform sustained writes of 100 GB\slash sec, a benchmark demonstrated that year by ASCI Purple.

Since 2005, the computational abilities of the most powerful supercomputer have increased by two orders of magnitude, from 100 teraflops to nearly 10 petaflops(taking K, June 2011’s \#1 as the example). At the same time, the total I\slash O throughput for these systems, measured in the total theoretical bandwidth for writes(the sum of the throughputs to each object storage target(OST) in the parallel file system) has remained nearly static. As this is being written, K writes across 864 OSTs to give realized I\slash O throughput of 96 GB\slash s \cite{Sumimoto2011} using the benchmarking software IOR \cite{Hedges2005}. 

This I\slash O plateau is not only observed with K; a 2008 paper \cite{Fahey2008} gives the results of very detailed testing of a Lustre based file system on a leadership scale computer, Jaguar of ORNL, a Cray XT. The system could perform computations in excess of 250 teraflops at the time, and the theoretical peak performance of the I\slash O system was 72 GB\slash s. The authors used MPI-IO based benchmarks to measure the actual performance of the filesystem and found that less than half of the theoretical peak could be achieved by such an approach. The poor actualized performance of parallel file systems with common I\slash O libraries is explored later in this report.

This report aims to make apparent the impending I\slash O challenges that will be faced by high performance computing systems. Modern scientific software development programs aim to create tools that will be in use for decades.  Necessarily, we must take into consideration the hardware upon which such tools will be executed. By introducing ideas taken from recent research in I\slash O into our software, we can ensure that it scales into the petaflop level and beyond. 

Furthermore, it is inevitable that our software will be run on a diverse selection of hardware with varying filesystems and performance characteristics. It is thus worthwhile to explore ways through which we can minimize the necessary effort to port the software across platforms. Recent research has been directed at this goal and should be brought to attention. 

\section{I\slash O in the High Performance Environment}
\label{iointhehighperformanceenvironment}

At scale, file system I\slash O is primarily used for application checkpointing. Writing to the filesystem dominates reading from it by a factor of 5 \cite{Hedges2005}. As noted in \cite{Moody2010}, the increases in computational performance of HPC systems can be to a large degree attributed to the use of more hardware components. As the complexity of the system increases, the average time to failure falls. Large scale modern machines, such as a 100,000 node BlueGene\slash L at LLNL have a failure once every 8 hours. In the future, at exascale, machines may have failure rates on the order of minutes. Furthermore, as the number of nodes increases, the amount of data that needs to be written with every checkpoint grows exponentially. Thus we have a need for increasing checkpoint frequency as well as more data to be written at each checkpoint.

\section{Data Intensive Applications and Scaling I\slash O}
\label{dataintensiveapplicationsandscalingio}

Data intensive applications have been enabled by the availability of powerful parallel file systems and parallel I\slash O libraries which enable applications to write out many terabytes of data at rates measured at up to 100 GB\slash s. However, as noted in \cite{Logan2010}, data intensive applications are straining the capabilities of even the most powerful of file systems. This problem will be even greater in the future as the number of compute cores in large scale systems will continue to rise at a rate much faster than the number of striped file targets in the parallel file system, primarily because of the expense in I\slash O hardware. Furthermore, parallel file systems rely on centralized metadata storage which can become overwhelmed by the network traffic required to communicate with many thousands of I\slash O clients. 

Optimizations to the parallel I\slash O system are considered difficult. These data intensive scientific codes access the I\slash O subsystem in ways that are not particularly amenable to parallelization. Frequently, these accesses take the pattern of many small, noncontiguous accesses of a single large file. As noted in \cite{Nisar}, this produces effective serialization of accesses as file locking semantics are required to maintain consistency and the guarantee of correctness. Parallel I\slash O frequently consists of many small atomic accesses. 

\cite{Fahey2008}, after a study on the I\slash O needs of developers of extreme scale applications, found that the programmer must increasingly become familiar with the intricacies of the I\slash O subsystem in order for their application to continue to scale. It is no longer sufficient to expect the parallel I\slash O libraries to continue working as they have in the past and provide sufficient performance with merely the addition of new hardware on the storage side. The scaling of I\slash O is currently at an impasse. One cannot continue with the ways of the past and expect the system to work as the number of cores in a computer system scales into the millions. One temporary, and necessary, stopgap in scaling I\slash O is data staging, further discussed below. This approach, now common, forwards the data from the compute nodes to a number of processes designated for I\slash O. In this way, network traffic between the clients and the file server is reduced and greater performance can be demonstrated.

\section{I\slash O as it Relates to Modern Software Development Programs}
\label{ioasitrelatestocreate}

Modern scientific software development programs  that seek to develop codes that will remain in production for decades must address the scalability problem. For our purposes, one can concentrate optimizing I\slash O in homogeneous, secure computing environments, which are the most common in leadership scale computing today. The leadership scale computers of today will within years become commonplace and the I\slash O challenges currently faced by the developers of the largest computers will need to be addressed by all. 

\pagebreak 

\chapter{Contemporary Approaches to Optimizing HPC I\slash O}
\label{contemporaryapproachestooptimizinghpcio}

\section{Data Staging}
\label{datastaging}

\subsection{Introduction}
\label{introduction}

The problem of I\slash O in the high performance environment is well recognized. Previous work has been done to address it on a variety of fronts. Important to the discussion is \cite{Fahey2008} which describes detailed tests run to analyze the performance of a Lustre parallel file system on Jaguar of ORNL. It introduces the concept of subsetting, derived from the observation that higher performance can be had by only using a fraction of the total number of processes to interact with the file servers. This can be attributed to less contention in communicating with the OSTs and that, generally, fewer, larger interactions with the file system are better than many smaller ones. An additional contribution of this paper is that it compares the performance of MPI-IO to HDF--5, an I\slash O framework that provides a higher level of abstraction, and the paper finds that write performance between the two can be comparable. 

These findings can be generalized into the notion of data staging. Rather than having all client nodes directly interacting with the file system servers, they forward their data and I\slash O requests to intermediary processes. Major work has been done in this area by the Center for Ultra-scale Computing and Information Security at Northwestern University and Georgia Tech’s Center for Experimental Research in Computer Systems (CERCS).

\subsection{Delegate Caching}
\label{delegatecaching}

\cite{Nisar} has developed a way to use what they term I\slash O delegates as intermediaries between the processes and the compute nodes. They observe that the BlueGene architecture has built in a layer between the compute nodes and the I\slash O servers through which requests can be funneled and draw inspiration from this in creating their prototype. Their system treats independent I\slash O the same as collective I\slash O, and they find that by using it, independent I\slash O can reach double the write performance of native collective I\slash O, an impressive achievement as collective operations have been traditionally better performing than independent I\slash O. 

Her work addresses the problem of serialization, common in independent I\slash O operations. As several processes contend to write to the same file, the file system must lock it and contending processes must be queued. By reducing the number of clients, there are fewer file locks. Furthermore, they implement a caching mechanism that reduces the number of I\slash O operations required by coupling smaller requests into a few larger ones.

\subsection{Flexible I\slash O Forwarding}
\label{flexibleioforwarding}

\cite{Ali2009} again reiterates the decline in I\slash O throughput compared to computational ability. They note that an increasing number of applications are becoming I\slash O bound and in the future this number will inevitably grow. Furthermore they state that the main difficulty in contemporary I\slash O research is to achieve maximum throughputs for existing architectures while also scaling up the application. One major problem in doing this is that the major modern parallel file systems, e.g. Lustre, PVFS, GPFS and PanFS were developed for smaller systems than what they are currently being used for. They also discuss the various levels at which enhancements to the I\slash O subsystem could be made, e.g., at the level of the parallel file system, at the level of the I\slash O library(generally ROMIO), or at the level between the two, the I\slash O forwarding layer, for which they also draw inspiration from the BlueGene systems. The forwarding they present would be transparent to the application and I\slash O libraries and work on multiple architectures, file systems and interconnects. 

\subsection{DataTaps and I\slash O Graphs}
\label{datatapsandiographs}

\cite{Widener2010a} looks at the use of DataTaps to add some structure to data which enable the data to be decoupled from the application in space and time. This header information is extremely lightweight, thus DataTaps can also be used as a substitute for typed I\slash O libraries such as HDF5 and NetCDF without their characteristic performance penalty. A smaller number of DataTap servers get the structured data off the compute nodes. From the DataTap servers, IO graphs are used to route the data to disk. 

\subsection{Irregular Accesses to Noncontiguous File Sections}
\label{irregularaccessestononcontiguousfilesections}

Scientific applications are not able reach I\slash O throughputs near the theoretical peak bandwidth of the I\slash O subsystems. This is because scientific applications do not typically make I\slash O requests in the optimal fashion. While some have been able to utilize collective I\slash O operations, as \cite{Nisar} points out, more and more are behaving erratically and requiring independent I\slash O which has been traditionally provided less throughput than collective I\slash O. 

\cite{Logan2010} directs its attention to the problem that scientific codes do not make best use of available I\slash O capacity. He says this is because scientific codes make I\slash O requests to a large number of non contiguous regions, with each request being of high latency. He states that most research in parallel I\slash O has been in attempting to optimize such requests. He argues that one can sidestep the issue by discarding the view of a file as a linear series of bytes and rather looking at data to be output as unrelated objects.

Lustre is the most common parallel file system for the largest supercomputer systems today, and it services 15 of the world’s top 30 computers(as of July 2011). As evidenced in \cite{Fahey2008}, MPI-IO is poorly supported by Lustre, and \cite{Logan2010} discusses the reasons for this in a large section of the paper. The primary problems, he claims, are that Lustre primarily works within the POSIX I\slash O view and thus poorly supports possible parallel optimizations. He demonstrates that the assumption that large, non-contiguous I\slash O requests are not necessarily the best way to attain performance but can rather lead to very poor throughputs. Also problematic for parallelization is that Lustre strictly enforces file consistency semantics, i.e. it locks files to protect from concurrent writes to the same region. These semantics limit performance and can be worked around by instead viewing the data to be output as unrelated objects.

\subsection{Interval I\slash O}
\label{intervalio}

\cite{Logan2010} is a dissertation that develops what they term Interval I\slash O, another method to add an intermediary layer between the computation and I\slash O servers. Primarily designed for multiple processes writing the same file, Interval I\slash O intelligently partitions a file into intervals using application access patterns. The benefits are similar to that described in \cite{Nisar} in that it reduces contention and serialization, especially for atomic operations. 

One important contribution of this work is that it argues that the view of a file as a flat contiguous series of bits is one of the most important contributors to poor parallel I\slash O performance. Rather, one should consider the data to be written out as discrete objects. If one were to do this, less contention would be had and greater parallelization made available. 

\subsection{Conclusion}
\label{conclusion}

The works above have in common the agreement that some intermediary layer must be added between the computational nodes and the I\slash O servers for continuing scalability. They agree that contemporary parallel file systems will be a bottleneck as the number of compute cores continues to scale. These data staging techniques presented are similar in that they all use a subset of processes to communicate with the file servers in order to reduce contention and communication. They differ in their respective implementations of that idea, specifically the level at which the data staging functionality is added. The level of implementation can be kernel space or user space, the implementation can either provide an application access to an API or it can be transparent to the application and forward the calls automatically. \cite{Logan2010}, a dissertation dated December 2010, provides a discussion of the differences in the various approaches to performing data staging. 

Data staging, however, is only a partial, and temporary, solution to the problem. The number of I\slash O delegated processes must increase along with the number of computational cores. As long as a central metadata server is required, there will be write contention. It would require a radical departure from the reigning model address this fundamental problem. Such an idea is reported below in chapter  \ref{promisingiodevelopmentsforthefuture} 

\pagebreak 

\section{Decoupling I\slash O From The Application Core}
\label{decouplingiofromtheapplicationcore}

\subsection{Introduction}
\label{introduction}

An important fact about production scientific codes is that they have a long lifespan. Once they have been thoroughly debugged, any modifications to the code must be accompanied by another review to ensure correctness. While the code may fundamentally stay the same over decades, the systems upon which it runs may change dramatically. Different institutions may also run the same code on different hardware organizations. 

One problem in the portability of code is that performance may vary dramatically between different parallel file systems. It is understood that, e.g., Lustre performs especially poorly with the MPI-IO library, and \cite{Logan2010} addresses a significant portion of his dissertation in the attempt to understand why. Other file systems handle I\slash O libraries with varying amounts of competency. By changing the I\slash O fundamentals of an application, it is possible to port it to computers with other I\slash O subsystems, but doing this within the source is burdensome and costly in man hours and requires maintaining multiple branches of a code. 

One can decouple I\slash O from the application’s core compute functionality by using abstraction. One makes calls to a middleware library which then maps the I\slash O to the particularity of whatever file systems the application happens to be running on. This decoupling is primarily for human convenience. This abstraction is present in the ADIOS I\slash O middleware. 

\cite{Widener2010a} discusses two more notions of decoupling I\slash O. One can decouple I\slash O applications temporally and spatially. Spatial decoupling is essentially data staging, i.e., having I\slash O requests forwarded from the compute nodes to nodes delegated for I\slash O. Temporal decoupling is similar to caching, combining multiple small I\slash O requests into fewer larger ones. Both notions are also present in \cite{Nisar} which uses delegates to decouple I\slash O spatially and a caching mechanism for decoupling I\slash O temporally.

\subsection{ADIOS}
\label{adios}

Researchers at Georgia Tech have developed a middleware system to enable applications to change the underlying I\slash O operations without making changes to the application source \cite{Lofstead2008}. One can program the application with high level I\slash O calls derived from POSIX and then use a configurable XML file to map those calls to what is optimal for the underlying hardware. This API, named ADIOS(Adaptable IO Systems) has been tested on leadership scale machines such as Jaguar and has achieved I\slash O throughputs comparable to that of lower level direct MPI-IO based benchmarks. 

\subsection{Exploiting Latent I\slash O Asynchrony}
\label{exploitinglatentioasynchrony}

\cite{Widener2010a} also addresses the need to decouple I\slash O from the application, and they look at the tolerance that an application has in decoupling I\slash O both temporally and spatially from the core computation. They note that the tolerance for asynchrony is a property of HPC codes that hasn’t been fully explored, but one can achieve significant performance improvements by overlapping computation with I\slash O. 

They use DataTaps to add headers, structuring the data, enabling it to be decoupled in space and time. DataTap servers pull the data from the DataTap clients. This is done during periods of relatively low activity within the interconnect to avoid interfering with application performance on the computational side. 

IOgraphs buffer the data, send it to special I\slash O nodes, and eventually write it to disk in structured form. Metabots further decouple I\slash O temporally by transforming the structured data on disk into a more regular form. This operation can be done at anytime spare CPU cycles are available. 

This system integrates with the ADIOS API to achieve another level of decoupling, that of indirection. It has been tested on large scale systems at Oak Ridge with applications such as GTC, a particle based fusion simulation that scales to hundreds of thousands of cores and must output tens of terabytes of data with each checkpoint. They describe that GTC can have bursts of output that exceed the bandwidth of the I\slash O subsystem, but with proper use of IOgraph buffering this can be overcome, allowing GTC to begin the next computation while the data is being written to disk. Their results also indicate that DataTaps can structure data at near peak bandwidth.

\subsection{Overlapping I\slash O with Iterative Solvers}
\label{overlappingiowithiterativesolvers}

\cite{Fu2010} also explores decoupling the I\slash O from the application in the context of an iterative solver. Rather than waiting on the I\slash O for each iteration to complete before beginning the next, they decouple the I\slash O temporally and move it to I\slash O nodes while continuing on iterating the solver. The result is perceived I\slash O bandwidth of the sum of the memory speed to the I\slash O nodes, or 21 TB\slash s in the example application they give.

\subsection{Conclusion}
\label{conclusion}

It is strongly recommended that developers of applications that may be used on varying hardware consider adapting such an approach to decrease the future amount of time required in maintaining the source code. By using such a mechanism, one can better future proof one’s code and allow the application to be adapted to yet to be developed file systems. Furthermore, by overlapping I\slash O with core computational functionality, one can achieve much higher bandwidth, both realized and perceived by the application.

\pagebreak 

\chapter{Promising I\slash O Developments for the Future}
\label{promisingiodevelopmentsforthefuture}

\section{Introduction}
\label{introduction}

\cite{Ali2009a}, a dissertation dated 2009, states that the problem in I\slash O scalability is worsened by contemporary parallel file systems and that the central server paradigm cannot handle the I\slash O requirements of modern data intensive applications. Furthermore, scientific applications increasingly depend on data stored in remote locations. Given examples of applications dependent upon dispersed data storage include the Large Hadron Collider, which generates data on the order of petabytes and because of the collaborative nature of the project, this data must be shared across wide area networks. 

These developments, he argues, call for a reevaluation of the current approach to I\slash O, one that ensures the scalability of applications to millions of cores and beyond and takes into account the often distributed environment in which the applications requiring the most compute capability will operate.

\cite{Raicu2011}, a very recent paper dated June 2011, states that the network storage based paradigm will no longer service exascale computers with billions of computational threads and that, if it were attempted, there would be a performance collapse. The mean time to failure would fall below the time required to checkpoint, and concurrent communication from clients would overwhelm the metadata server. Rather, they propose that advances in storage technology will enable distributed metadata services on the compute nodes. 

They point out that the gap between computational capacity and I\slash O bandwidth has seen a 10x increase over the past decade. They argue that the most significant barrier to building exascale computers will be the I\slash O subsystem. Common I\slash O operations, when performed by millions of concurrent processes, will take more time than the mean time to failure. Even the action of booting the machine is projected to take 7 hours. 

They argue that distributed file systems should exist in tangent with classical parallel file systems and be used for the jobs requiring the full computational capacity of the machine. Every node would be equipped with solid state storage and use the multicore capabilities to participate in the management of both metadata and data. They note that this approach was previously unfeasible because of the high rate of failure in mechanical magnetic storage devices, but new developments in solid state storage may attain reliability sufficient. The authors are working on a prototype of such a distributed storage system named FusionFS.

\section{Questioning the Checkpointing Paradigm}
\label{questioningthecheckpointingparadigm}

As noted earlier, checkpointing dominates the I\slash O for large applications, and writes can dominate reads by a factor of 5. As the number of compute nodes scale, the total amount of system memory increases and thus the amount of data that needs to be dumped at each checkpoint is greater. Furthermore, as the hardware complexity of computer systems increase, the mean time to failure falls, and the frequency of checkpointing must rise. 

Some researchers have sought to address the scalable I\slash O problem by questioning this checkpointing paradigm. Currently the practice is to have each compute node dump the same large subset of the data it holds in memory to disk at every checkpoint. The data written must be sufficient to restart the application in the event of a failure.

\cite{Moody2010} introduces a multi level checkpointing system. They analyze the frequency of different types of failure in an HPC system and observe that the most common types of failures do not require a full heavyweight checkpoint. Rather, the most common failures only disable a node or two at one time and that failures of multiple nodes happen predictably. They report that 85\% of failures disable at most one compute node in the cluster. By writing checkpoint data to RAM, flash memory, or scratch disk space on the compute nodes, these failures can be protected for while achieving effective checkpointing speeds greater than two orders of magnitude faster than the existing model. The fewer, less frequent heavyweight checkpoints protect against multinode failures.

One drawback of this approach is that it requires storage on the compute nodes. This would either mean one would use a portion of the ram or have some kind of non-volatile storage available on each node. \cite{Raicu2011} argues that systems at the exascale will indeed have such components and will have to make use of them for file system data management. Some combination of multi level checkpointing with distributed parallel file systems at the node level may be a viable solution for future scalability. 

Diskless checkpointing is the use of volatile memory to add levels of redundancy by storing just enough information to allow for the recomputation of the original data in case of a node failure. Analogous to RAID technology that is instrumental in affording redundancy of disk drives, diskless checkpointing stores parity bytes in the memory of other nodes that, when added to the data held by other nodes, allows for data recovery when one node goes down. While this would eliminate the need to store checkpointing data to the parallel file system, it has traditionally been too computationally expensive to consider. However, novel architectures are becoming more prevalent in HPC environments and heterogeneous compute nodes with highly parallel accelerator cards are now common. One can leverage the highly parallel nature of e.g. GPUs to accelerate Reed-Solomon computation of the parity information and make diskless checkpointing practical. 

\cite{Gomez}, to be published in the November 2011 Supercomputing proceedings, develops a very high frequency low overhead checkpointing system using heterogeneous components. Noting also that post petascale systems will have very small mean times to failures and extremely large amounts of data to write out(hundreds of teraflops to petabytes), they argue that it is essential to develop a novel approach to checkpointing. They discuss the most modern supercomputing systems, noting that they have solid state memory or some other type of non volatile storage space on the compute node.

While soft errors can always be recovered by the data stored on the compute node, even hard errors, i.e., the full node going down, can be recovered from if they make use of partner nodes to replicate erasure codes(typical RAID-like parity byte encoding). This requires an extra transfer of the code but is more efficient than using the central parallel file system to protect against hard failures. This approach does require that one takes into consideration the topology(spatial arrangement) of the nodes, as one doesn’t want to store parity bytes on partner nodes that fail predictably together. 

They build upon the multi level checkpointing system discussed above, but theirs contains three levels. The first level protects against soft errors by storing checkpoint data locally on the node. The second level uses heterogeneous Reed-Solomon coding to create parity bytes and stores them on partner nodes. This can be used to recover from the most common type of hard failures. The third level and least frequent writes checkpoint data out to the parallel file system. This protects against catastrophic system failures. The primary focus of their paper is the second level, or the use of heterogeneous architecture to protect against limited hard failure.

One more interesting point they cover is that as the fabrication process decreases in scale, e.g., from 90 nm to 16 nm, the rate of soft errors increases significantly. This observation leads to a discussion of pattern prediction in failures which enable intelligently developed redundancy systems. 

\section{Are Parallel File Systems Really Parallel?}
\label{areparallelfilesystemsreallyparallel}

\cite{Ali2009} notes that modern parallel file systems were really developed for smaller supercomputing systems than what they are currently being used for. \cite{Nisar} states that achieving scalability is simply not possible without a fundamental change to the premises upon which they operate. What one considers as parallel distributed file systems are in fact not really distributed, but rather depend on a central metadata server to drive the requests. Communication with this server can become saturated with too many client processes and performance can fall dramatically. 

The performance improvements that data staging brings are because what we commonly understand to be distributed file systems are fundamentally an outgrowth of traditional serial file systems. However, very recent developments in new hardware, some researchers are questioning whether it would be appropriate to implement a fully distributed file system, one that is no longer dependent upon a central metadata server. 

Both \cite{Ali2009a} and \cite{Raicu2011} develop models of distributed parallel file systems that are not dependent upon a central metadata server. \cite{Ali2009a} is applicable to large scale collaborative scientific computing that relies on resources distributed across wide area networks. \cite{Raicu2011} advocates a distributed file system using contemporary developments in solid state storage hardware which offers resiliency levels previously unmet by magnetic disk drives. With such resiliency, one can have each compute node participating in metadata and data management rather than having it managed by a redundant central metadata server. 

\section{In Situ Analytics}
\label{insituanalytics}

Petascale applications generate data of such tremendous scale(on the order of tens of terabytes per checkpoint or iteration) that the post processing of it becomes extremely time consuming. Data output has reached such a level that, according to Dr. Karsten Schwan of Georgia Tech, if one does not perform some kind of analysis on it while in transit, it is unlikely that one can fully process it. In an application one always makes the choice of what subset of data to output. With in situ analytics, one can intelligently discover interesting parts of data on the fly while the application is running. Using such techniques enable us to stop or correct an erroneous run, manipulate or refine a mesh, or provide input to guide an application to a more stable solution. 

The in situ analytics of petascale applications are being studied by a number of labs. One of the most prominent is the visual study of combustion at Sandia \cite{Yu2009}. This is case study of using in situ visualization using the direct numerical simulation of turbulence at small scale. While scientists have traditionally tried to dump as much raw data as they possibly can for post processing analysis, in situ analytics attempts to tame the flood of data as it is in transit and prepare it for visualization while the application is being run. 

At the petascale, it is very difficult to post process data because one does not typically have petabytes of data storage available and the transfer of such amounts of data would be prohibitively expensive. While one can only transfer and examine certain time steps, this defeats the point of high resolution simulations in the first place. \cite{Yu2009} argues that the most effective way to process the great amount of data is directly on each compute node in situ. By doing this we save both the cost of transfer time of data and the time required to post process it. They address the unique challenges of doing visualization in situ, and they find that visualization calculations only take an acceptably small fraction of the total supercomputing resources. 

\pagebreak 

\chapter{Conclusion}
\label{conclusion}

\section{Levels of Indirection and Planning for the Future}
\label{levelsofindirectionandplanningforthefuture}

“All problems in computer science can be solved by another level of indirection” states the famous aphorism in computing. For new development projects, we must ensure that our applications scale into the petaflop level and are easily portable to other hardware architectures by considering the research presented in this report. 

The ADIOS API, currently being maintained by Georgia Tech and Sandia National Labs provides a simple, POSIX based interface for cross platform I\slash O programming. By modifying and tweaking an accompanying XML file, we can get high performance using a variety of I\slash O libraries which would make our applications more portable to other file system architectures. 

By integrating such an approach with advances in data staging, we can make sure that our application can scale into the tens or hundreds of thousands of cores without overwhelming the central servers in a parallel file system. By exploiting latent asynchrony, or the tolerance for an application to overlap computation and ancillary I\slash O operations, we can achieve much higher perceived throughputs to the I\slash O subsystem. 

Lastly, as we are developing applications that should remain in use for the next few decades, we should at least consider the new research in both multi level checkpointing and the potential for disruptive new revolutions in the fundamental architecture of I\slash O subsystems. 

\bibliography{Ph.D.Research-IOPaperSummer2011}

\begin{thebibliography}{10}

\bibitem{Ali2009a}
Nawab Ali.
\newblock {\em {Rethinking I/O in High-Performance Computing Environments}}.
\newblock PhD thesis, The Ohio State University, 2009.

\bibitem{Ali2009}
Nawab Ali, Philip Carns, Kamil Iskra, Dries Kimpe, Samuel Lang, Robert Latham,
  Robert Ross, Lee Ward, and P.~Sadayappan.
\newblock {Scalable I/O forwarding framework for high-performance computing
  systems}.
\newblock In {\em 2009 IEEE International Conference on Cluster Computing and
  Workshops}, pages 1--10. IEEE, 2009.

\bibitem{Fahey2008}
M.~Fahey, J.~Larkin, and J.~Adams.
\newblock {I/O performance on a massively parallel Cray XT3/XT4}.
\newblock In {\em Parallel and Distributed Processing, 2008. IPDPS 2008. IEEE
  International Symposium on}, pages 1--12, 2008.

\bibitem{Fu2010}
J.~Fu, N.~Liu, O.~Sahni, K.~E. Jansen, M.~S. Shephard, and C.~D. Carothers.
\newblock {Scalable parallel I/O alternatives for massively parallel
  partitioned solver systems}.
\newblock {\em 2010 IEEE International Symposium on Parallel \& Distributed
  Processing, Workshops and Phd Forum (IPDPSW)}, pages 1--8, April 2010.

\bibitem{Gomez}
L.~B. Gomez, Dimitri Komatitsch, N.~Maruyama, S.~Tsuboi, F.~Cappello, and
  S.~Matsuoka.
\newblock {FTI: high performance Fault Tolerance Interface for hybrid systems}.
\newblock In {\em Supercomputing 2011}.

\bibitem{Hedges2005}
Richard Hedges, Bill Loewe, T.~McLarty, and Chris Morrone.
\newblock {Parallel File System Testing for the Lunatic Fringe: the care and
  feeding of restless I/O Power Users}.
\newblock In {\em Proceedings of the 22nd IEEE/13th NASA Goddard Conference on
  Mass Storage Systems and Technologies}, number Msst. Published by the IEEE
  Computer Society, 2005.

\bibitem{Lofstead2008}
J.F. Lofstead, S.~Klasky, K.~Schwan, Norbert Podhorszki, and Chen Jin.
\newblock {Flexible io and integration for scientific codes through the
  adaptable io system (adios)}.
\newblock In {\em Proceedings of the 6th international workshop on Challenges
  of large applications in distributed environments}, pages 15--24. ACM, 2008.

\bibitem{Logan2010}
Jeremy Logan.
\newblock {\em {Improving Parallel I/O Performance Using Interval I/O}}.
\newblock PhD thesis, The University of Maine, 2010.

\bibitem{Moody2010}
Adam Moody, Greg Bronevetsky, Kathryn Mohror, and Bronis R~De Supinski.
\newblock {Design , Modeling , and Evaluation of a Scalable Multi-level
  Checkpointing System}.
\newblock In {\em Supercomputing 2010}, number November, 2010.

\bibitem{Nisar}
Arifa Nisar, Wei-keng Liao, and Alok Choudhary.
\newblock {Delegation-based I/O Mechanism for High Performance Computing
  Systems}.
\newblock In {\em IPDPS 2011}, 2011.

\bibitem{Raicu2011}
Ioan Raicu, I.T. Foster, and Pete Beckman.
\newblock {Making a case for distributed file systems at Exascale}.
\newblock In {\em LSAP 2011}, pages 11--18. ACM, 2011.

\bibitem{Sumimoto2011}
Shinji Sumimoto.
\newblock {An Overview of Fujitsu's Lustre Based File System}.
\newblock Technical report, Fujitsu, 2011.

\bibitem{Widener2010a}
P.~Widener, M.~Wolf, H.~Abbasi, S.~McManus, M.~Payne, M.~Barrick,
  J.~Pulikottil, P.~Bridges, and K.~Schwan.
\newblock {Exploiting Latent I/O Asynchrony in Petascale Science Applications}.
\newblock {\em International Journal of High Performance Computing
  Applications}, 25(2):161--179, May 2010.

\bibitem{Yu2009}
Hongfeng Yu, J.H. Chen, Chaoli Wang, K.~Ma, and R.W. Grout.
\newblock {A Study of In-Situ Visualization for Petascale Combustion
  Simulations}.
\newblock Technical report, 2009.

\end{thebibliography}
\end{document}